\documentclass[aps,preprint]{revtex4}%
\usepackage{amsfonts}
\usepackage{amsmath}
\usepackage{amssymb}
\usepackage{graphicx}%
\begin{document}
\title{Exact Pointer Properties for Quantum System Projector Measurements with
Application to Weak Measurements and Their Accuracy }
\author{A. D. Parks and J. E. Gray}
\affiliation{Electromagnetic and Sensor Systems Department, Naval Surface Warfare Center,
Dahlgren, VA 22448, USA}

\begin{abstract}
Exact pointer states are obtained for projection operator measurements
performed upon pre-selected (PS) and upon pre- and post-selected (PPS) quantum
systems. These states are used to provide simple exact expressions for both
the pointer spatial probability distribution profiles and the mean values of
arbitrary pointer observables associated with PS and PPS projection operator
measurements that are valid for any strength of the interaction which couples
a measurement pointer to the quantum system. These profiles and mean values
are compared in order to identify the effects of post-selection upon projector
measurement pointers. As a special case, these mean value results are applied
to the weak measurement regime - yielding PS and PPS mean value expressions
which are valid for any operator (projector or non-projector). Measurement
sensitivities which are useful for estimating weak measurement accuracies for
PS and PPS systems are also obtained and discussed.

\end{abstract}
\date{October 7, 2010}
\startpage{1}
\endpage{102}
\maketitle

\section{Introduction}

The pointer of a measurement apparatus is fundamental to quantum measurement
theory because the values of measured observables are determined from its
properties (e.g. from the mean values for the pointer position and momentum
operators $\widehat{q}$ and $\widehat{p}$, respectively). Understanding these
properties has become more important in recent years - in large part due to
the increased interest in the theory of weak measurements of pre- and
post-selected (PPS) quantum systems and weak value theory. Because of this
growing interest the practical value of estimating the associated measurement
sensitivities has also become important from both the experimental and device
engineering perspectives.

The use of PPS techniques for controlling and manipulating quantum systems was
introduced by Schr\"{o}dinger more than $75$ years ago \cite{S1,S2}. Since
then PPS techniques have found utility in such diverse areas of study as
quantum system-environment interactions, e.g. \cite{R} ; the quantum eraser,
e.g. \cite{K} ; and Pancharatnam phase, e.g. \cite{H,W}. One especially
fertile area of application of PPS theory is the time symmetric reformulation
of quantum mechanics developed by Aharonov, Bergman and Lebowitz \cite{ABL}
and the closely related notion of the weak value of a quantum mechanical
observable, e.g. \cite{A1,A2,A3}.

The weak value $A_{w}$ of a quantum mechanical observable $A$ is the
statistical result of a standard measurement procedure performed upon a PPS
ensemble of quantum systems when the interaction between the measurement
apparatus and each system is sufficiently weak, i.e. when it is a weak
measurement. Unlike a standard strong measurement of $A$ performed upon a
prepared, i.e. pre-selected (PS), ensemble which significantly disturbs the
measured system and yields the mean value of the associated operator
$\widehat{A}$ as the measured value of $A$, a weak measurement of $A$
performed upon a PPS system does not appreciably disturb the quantum system
and yields $A_{w}$ as the observable's measured value. The peculiar nature of
the virtually undisturbed quantum reality that exists between the boundaries
defined by the PPS states is revealed by the eccentric characteristics of
$A_{w}$, namely that $A_{w}$ can be complex valued and that $\operatorname{Re}%
A_{w}$ can lie far outside the eigenvalue spectral limits of $\widehat{A}$.
While the interpretation of weak values remains somewhat controversial,
experiments have verified several of the interesting unusual properties
predicted by weak value theory \cite{RSH,PCS,WSZ,HK,DSJ}. The theory has also
been applied to the theoretical and experimental resolution of such quantum
paradoxes as "the quantum box problem" \cite{AV,V,RLS} and "Hardy's Paradox"
\cite{H1,H2,ABP,LS,YYK}.

As is well known, projection operators are an important part of the general
mathematical formalism of quantum mechanics. There has been a recent increased
interest in these operators because the measurement and interpretation of
their weak values have played a central role in the theoretical and
experimental resolutions of the quantum box problem and Hardy's paradox. More
recently projector weak values have also been exploited in experimental
observations of dynamical non-locality induced effects \cite{TAC,SP}.

Projection operators are also interesting because their idempotent property
can be used to provide simple exact descriptions of pointers resulting from
projection operator measurements. Specifically, when an instantaneous
measurement is performed upon a quantum system to determine the value of a
projection operator $\widehat{A}$, the associated von Neumann measurement
interaction operator $e^{-\frac{i}{\hbar}\gamma\widehat{A}\widehat{p}}$ is
easily shown (using the series expansion of the measurement interaction
operator and the generalized idempotent property $\widehat{A}^{n}=\widehat
{A},n\geq1$) to be given exactly by%
\begin{equation}
e^{-\frac{i}{\hbar}\gamma\widehat{A}\widehat{p}}=\widehat{1}-\widehat
{A}+\widehat{A}\widehat{S}. \label{1}%
\end{equation}
Here $\gamma$ is the measurement interaction strength and $\widehat{S}\equiv
e^{-\frac{i}{\hbar}\gamma\widehat{p}}$ is the pointer position translation
operator defined by its action $\left\langle q\right\vert \widehat
{S}\left\vert \phi\right\rangle \equiv\phi\left(  q-\gamma\right)  $ upon the
pointer state $\left\vert \phi\right\rangle $ (it is assumed hereinafter that
the commutator $\left[  \widehat{A},\widehat{S}\right]  =0$).

The primary objectives of this paper are to : (i) use eq.(\ref{1}) to provide
simple exact expressions for pointer spatial probability distribution profiles
and for the mean values of arbitrary pointer observables when projection
operator measurements of any interaction strength are performed upon PS and
upon PPS quantum systems (consequently, these results describe - without
approximation - the properties of such pointers over the entire range of
measurement interaction strengths - i.e. from the weak measurement regime to
the strong measurement regime) ; (ii) identify some of the effects induced
upon the pointer via state post-selection by comparing these PS and PPS
pointer results ; (iii) use these exact pointer results to obtain pointer mean
values for PS and PPS systems which are valid for any operator (i.e.,
projector or non-projector) in the weak measurement regime, i.e. when
$0<\gamma\ll1$ ; and (iv) obtain measurement sensitivity expressions which are
useful for estimating the accuracy of weak PS and PPS measurements of any such
operator. It is important to note that when direct comparisons between
quantities are made it is assumed that the interaction strength is fixed.

This paper is organized as follows : the next section is a development,
comparison, and discussion of the exact theories for pointers resulting from
projector operator measurements performed upon PS and PPS systems. Section III
is devoted to obtaining from the exact mean values approximate pointer
observable mean values that are valid for any operator in the weak measurement
regime - as well as to developing measurement sensitivities associated with
the weak measurement of any PS and PPS system observable. Concluding remarks
comprise the paper's final section.

\section{Exact Pointer Theories for Projector Measurements}

\subsection{PS Systems}

Consider the measurement at time $t$ of a time independent projector
$\widehat{A}$ performed upon a quantum system prepared in the normalized PS
state $\left\vert \psi\right\rangle $. Let the pointer of the measuring
apparatus be intitally in the pre-measurement state $\left\vert \phi
\right\rangle $. Then - from eq.(\ref{1}) - the exact normalized state
$\left\vert \Phi\right\rangle $ of the pointer immediately after the
measurement is%
\begin{equation}
\left\vert \Phi\right\rangle =e^{-\frac{i}{\hbar}\gamma\widehat{A}\widehat{p}%
}\left\vert \psi\right\rangle \left\vert \phi\right\rangle =\left(
\widehat{1}-\widehat{A}+\widehat{A}\widehat{S}\right)  \left\vert
\psi\right\rangle \left\vert \phi\right\rangle \label{2}%
\end{equation}
(the normalization of $\left\vert \Phi\right\rangle $ follows directly from
the fact that $\left(  \widehat{1}-\widehat{A}+\widehat{A}\widehat{S}\right)
^{-1}=\left(  \widehat{1}-\widehat{A}+\widehat{A}\widehat{S}\right)  ^{\dag
}=\left(  \widehat{1}-\widehat{A}+\widehat{A}\widehat{S}^{\dag}\right)  $).
The associated exact spatial probability distribution profile $\left\vert
\left\langle q\right\vert \left.  \Phi\right\rangle \right\vert ^{2}$ of the
pointer is given by%
\begin{equation}
\left\vert \left\langle q\right\vert \left.  \Phi\right\rangle \right\vert
^{2}=\left(  1-\left\langle \psi\right\vert \widehat{A}\left\vert
\psi\right\rangle \right)  \left\vert \left\langle q\right\vert \left.
\phi\right\rangle \right\vert ^{2}+\left\langle \psi\right\vert \widehat
{A}\left\vert \psi\right\rangle \left\vert \left\langle q\right\vert
\widehat{S}\left\vert \phi\right\rangle \right\vert ^{2}, \label{3}%
\end{equation}
which is simply the weighted sum of the distribution profiles for the
pre-measurement state $\left\vert \phi\right\rangle $ and $\widehat
{S}\left\vert \phi\right\rangle $ - the pre-measurement state translated by
$\gamma$. Observe that \emph{the idempotency of }$\widehat{A}$\emph{ precludes
the existence of an interference cross term} proportional to
$\operatorname{Re}\left\langle q\right\vert \left.  \phi\right\rangle ^{\ast
}\left\langle q\right\vert \widehat{S}\left\vert \phi\right\rangle $ in
eq.(\ref{3}) because the cross terms contain $\left\langle \psi\right\vert
\widehat{A}\left(  \widehat{1}-\widehat{A}\right)  \left\vert \psi
\right\rangle =\left\langle \psi\right\vert \left(  \widehat{A}-\widehat
{A}^{2}\right)  \left\vert \psi\right\rangle =\left\langle \psi\right\vert
\left(  \widehat{A}-\widehat{A}\right)  \left\vert \psi\right\rangle =0$ and
$\left\langle \psi\right\vert \left(  \widehat{1}-\widehat{A}\right)
\widehat{A}\left\vert \psi\right\rangle =0$ as factors.

If $\widehat{M}$ is the operator for any pointer observable $M$, then - from
eq.(\ref{2}) - the exact expression for the mean value of $M$ after a PS
measurement is readily found to be%
\begin{equation}
\left\langle \Phi\right\vert \widehat{M}\left\vert \Phi\right\rangle =\left(
1-\left\langle \psi\right\vert \widehat{A}\left\vert \psi\right\rangle
\right)  \left\langle \phi\right\vert \widehat{M}\left\vert \phi\right\rangle
+\left\langle \psi\right\vert \widehat{A}\left\vert \psi\right\rangle
\left\langle \phi\right\vert \widehat{S}^{\dag}\widehat{M}\widehat
{S}\left\vert \phi\right\rangle \label{4}%
\end{equation}
(it is assumed here - and hereinafter - that the commutator $\left[
\widehat{A},\widehat{M}\right]  =0$). This is the result anticipated from the
form of eq.(\ref{3}) - i.e. no interference cross term appears in eq.(\ref{4})
and it is simply the weighted sum of the pre-measurement mean value of $M$ and
its mean value relative to the $\gamma$ translated state $\widehat
{S}\left\vert \phi\right\rangle $. Note that $\left\langle \Phi\right\vert
\widehat{M}\left\vert \Phi\right\rangle =\left\langle \phi\right\vert
\widehat{M}\left\vert \phi\right\rangle $ when : (i) $\gamma=0$, i.e. no
measurement takes place ; or (ii) $\left\langle \psi\right\vert \widehat
{A}\left\vert \psi\right\rangle =0$. Also note that when $\widehat{M}%
=\widehat{p}$, then - since $\left[  \widehat{p},\widehat{S}\right]  =0$ -
$\left\langle \phi\right\vert \widehat{S}^{\dag}\widehat{p}\widehat
{S}\left\vert \phi\right\rangle =\left\langle \phi\right\vert \widehat
{p}\left\vert \phi\right\rangle $ and eq.(\ref{4}) becomes
\begin{equation}
\left\langle \Phi\right\vert \widehat{p}\left\vert \Phi\right\rangle
=\left\langle \phi\right\vert \widehat{p}\left\vert \phi\right\rangle .
\label{4a}%
\end{equation}
Thus the mean value of the pointer momentum is not changed by the projector
measurement of a PS system, i.e. \emph{pointer momentum is a constant of the
motion for projector measurements of PS systems} (in fact - pointer momentum
is a constant of the motion for both projector and non-projector measurements
of PS systems since $\left[  \widehat{p},e^{-\frac{i}{\hbar}\gamma\widehat
{A}\widehat{p}}\right]  =0$ so that $\left\langle \Phi\right\vert \widehat
{p}\left\vert \Phi\right\rangle =\left\langle \phi\right\vert \left\langle
\psi\right\vert e^{\frac{i}{\hbar}\gamma\widehat{A}\widehat{p}}\widehat
{p}e^{-\frac{i}{\hbar}\gamma\widehat{A}\widehat{p}}\left\vert \psi
\right\rangle \left\vert \phi\right\rangle =\left\langle \phi\right\vert
\widehat{p}\left\vert \phi\right\rangle $).

The fact that no interference cross terms appear in eq.(\ref{4}) is a useful
feature of PS system pointers. In particular, when $\widehat{M}=\widehat{q}$
and $\left\vert \phi\right\rangle $ is such that $\left\langle \phi\right\vert
\widehat{q}\left\vert \phi\right\rangle =0$ and $\left\langle \phi\right\vert
\widehat{S}^{\dag}\widehat{q}\widehat{S}\left\vert \phi\right\rangle =\gamma$
- e.g., when $\left\vert \left\langle q\right\vert \left.  \phi\right\rangle
\right\vert ^{2}$ is Gaussian with $0$ mean - then it is found from
eq.(\ref{4}) that%
\begin{equation}
\left\langle \Phi\right\vert \widehat{q}\left\vert \Phi\right\rangle
=\gamma\left\langle \psi\right\vert \widehat{A}\left\vert \psi\right\rangle .
\label{5}%
\end{equation}
Thus - in this case - \emph{if }$\gamma$\emph{ }(\emph{or }$\left\langle
\psi\right\vert \widehat{A}\left\vert \psi\right\rangle $)\emph{ is known,
then }$\left\langle \psi\right\vert \widehat{A}\left\vert \psi\right\rangle
$\emph{ }(\emph{or }$\gamma$)\emph{ can be determined directly from the
measurement pointer's mean position. }

\subsection{PPS Systems}

Now suppose that a measurement of projector $\widehat{A}$ is performed at time
$t$ upon a PPS system. If - as above - the pre-measurement pointer state is
$\left\vert \phi\right\rangle $, then the exact normalized pointer state
$\left\vert \Psi\right\rangle $ immediately after the post-selection
measurement is%
\begin{equation}
\left\vert \Psi\right\rangle =\frac{1}{N}\left(  \frac{\left\langle \psi
_{f}\right\vert \left.  \psi_{i}\right\rangle }{\left\vert \left\langle
\psi_{f}\right\vert \left.  \psi_{i}\right\rangle \right\vert }\right)
\left(  1-A_{w}+A_{w}\widehat{S}\right)  \left\vert \phi\right\rangle ,
\label{6}%
\end{equation}
where $\left\vert \psi_{i}\right\rangle $ and $\left\vert \psi_{f}%
\right\rangle $ are the normalized pre- and post-selected states at $t$,
respectively; $A_{w}$ is the weak value of $A$ at $t$ defined by%
\[
A_{w}\equiv\frac{\left\langle \psi_{f}\right\vert \widehat{A}\left\vert
\psi_{i}\right\rangle }{\left\langle \psi_{f}\right\vert \left.  \psi
_{i}\right\rangle },\text{ \ }\left\langle \psi_{f}\right.  \left\vert
\psi_{i}\right\rangle \neq0;
\]
and%
\[
N\equiv\left(  \left\vert N\right\vert ^{2}\right)  ^{\frac{1}{2}}%
=\sqrt{\left\vert 1-A_{w}\right\vert ^{2}+\left\vert A_{w}\right\vert
^{2}+2\operatorname{Re}\left[  A_{w}\left(  1-A_{w}^{\ast}\right)
\left\langle \phi\right\vert \widehat{S}\left\vert \phi\right\rangle \right]
}.
\]
Using the fact that \cite{AA}
\[
e^{i\chi}=\frac{\left\langle \psi_{f}\right\vert \left.  \psi_{i}\right\rangle
}{\left\vert \left\langle \psi_{f}\right\vert \left.  \psi_{i}\right\rangle
\right\vert },
\]
where $\chi$ is the Pancharatnam phase, enables eq.(\ref{6}) to be more
compactly written as%
\begin{equation}
\left\vert \Psi\right\rangle =\frac{e^{i\chi}}{N}\left(  1-A_{w}+A_{w}%
\widehat{S}\right)  \left\vert \phi\right\rangle . \label{7}%
\end{equation}
The associated exact spatial probability distribution profile $\left\vert
\left\langle q\right\vert \left.  \Psi\right\rangle \right\vert ^{2}$ of the
pointer is%
\begin{equation}
\left\vert \left\langle q\right\vert \left.  \Psi\right\rangle \right\vert
^{2}=\left(  \frac{1}{N^{2}}\right)  \left\{
\begin{array}
[c]{c}%
\left\vert 1-A_{w}\right\vert ^{2}\left\vert \left\langle q\right\vert \left.
\phi\right\rangle \right\vert ^{2}+\left\vert A_{w}\right\vert ^{2}\left\vert
\left\langle q\right\vert \widehat{S}\left\vert \phi\right\rangle \right\vert
^{2}+\\
2\operatorname{Re}\left[  A_{w}\left(  1-A_{w}^{\ast}\right)  \left\langle
q\right\vert \left.  \phi\right\rangle ^{\ast}\left\langle q\right\vert
\widehat{S}\left\vert \phi\right\rangle \right]
\end{array}
\right\}  . \label{8}%
\end{equation}

The effect of post-selection upon pointer states can be seen by comparing
eq.(\ref{7}) with eq.(\ref{2}). \emph{Eventhough the measurements are
generally not weak measurements}, it is interesting that -unlike projector
measurement pointer states for PS systems which depend upon $\widehat
{A}\left\vert \psi\right\rangle $ - \emph{projector measurement pointer states
for PPS systems explicitly depend upon the projector's weak value }$A_{w}$.
This - perhaps - is not surprising in light of the recent discussion in
\cite{AB} concerning von Neumann measurements and the associated ubiquitous
nature of weak values. It is also apparent from this comparison that
\emph{state post-selection is responsible for the presence of the Pancharatnam
phase factor }$e^{i\chi}$\emph{ in PPS pointer states}. This is an expected
natural consequence of state post-selection \cite{AA,S,TKN}.

As is the case for PS measurements, the pointer state distribution profiles
for PPS measurements are also weighted sums of the distribution profiles for
$\left\vert \phi\right\rangle $ and $\widehat{S}\left\vert \phi\right\rangle
$. However - unlike the PS case - \emph{the profile for PPS measurements
contains interference cross terms induced by state post-selection}.
Interference occurs because post-selection nullifies the idempotency of
$\widehat{A}$ by replacing $\widehat{A}\left\vert \psi\right\rangle $ with
$A_{w}$ - thereby allowing the cross terms to occur. More specifically -
unlike a PS measurement where the cross terms contain the vanishing
$\left\langle \psi\right\vert \widehat{A}\left(  \widehat{1}-\widehat
{A}\right)  \left\vert \psi\right\rangle $ and $\left\langle \psi\right\vert
\left(  \widehat{1}-\widehat{A}\right)  \widehat{A}\left\vert \psi
\right\rangle $ factors - the cross terms for PPS measurements contain
$A_{w}\left(  1-A_{w}^{\ast}\right)  $ and its complex conjugate as
non-vanishing factors.

As before, let $\widehat{M}$ be the operator for an arbitrary pointer
observable $M$. Using eq.(\ref{7}) it is found that the exact expression for
the mean value of $M$ after a PPS measurement has the form anticipated from
that of eq.(\ref{8}) :%
\begin{equation}
\left\langle \Psi\right\vert \widehat{M}\left\vert \Psi\right\rangle =\left(
\frac{1}{N^{2}}\right)  \left\{
\begin{array}
[c]{c}%
\left\vert 1-A_{w}\right\vert ^{2}\left\langle \phi\right\vert \widehat
{M}\left\vert \phi\right\rangle +\left\vert A_{w}\right\vert ^{2}\left\langle
\phi\right\vert \widehat{S}^{\dag}\widehat{M}\widehat{S}\left\vert
\phi\right\rangle +\\
2\operatorname{Re}\left[  A_{w}\left(  1-A_{w}^{\ast}\right)  \left\langle
\phi\right\vert \widehat{M}\widehat{S}\left\vert \phi\right\rangle \right]
\end{array}
\right\}  . \label{9}%
\end{equation}
Note that - similar to the PS case - for PPS measurements $\left\langle
\Psi\right\vert \widehat{M}\left\vert \Psi\right\rangle =\left\langle
\phi\right\vert \widehat{M}\left\vert \phi\right\rangle $ when : (i)
$\gamma=0$ ; or (ii) $A_{w}=0$. Also observe from eq.(\ref{9}) that when
$\widehat{M}=\widehat{p}$ and $A_{w}\neq0\neq\gamma$, then%
\[
\left\langle \Psi\right\vert \widehat{p}\left\vert \Psi\right\rangle =\left(
\frac{1}{N^{2}}\right)  \left\{
\begin{array}
[c]{c}%
\left(  \left\vert 1-A_{w}\right\vert ^{2}+\left\vert A_{w}\right\vert
^{2}\right)  \left\langle \phi\right\vert \widehat{p}\left\vert \phi
\right\rangle +\\
2\operatorname{Re}\left[  A_{w}\left(  1-A_{w}^{\ast}\right)  \left\langle
\phi\right\vert \widehat{p}\widehat{S}\left\vert \phi\right\rangle \right]
\end{array}
\right\}
\]
so that - unlike PS systems - measurements of PPS systems generally do change
the mean value of $p$. Thus, \emph{pointer momentum} \emph{is not a constant
of the motion for projector measurements of PPS systems} (this - in fact - is
also the case for non-projector measurements of PPS systems).

Because of the interference term in eq.(\ref{9}), pointer positions for PPS
systems are not as straightforwardly useful as those for PS systems for
measuring $\widehat{A}$ or $\gamma$. However, \emph{for the special case
}$\widehat{M}=\widehat{q}$\emph{, }$A_{w}=1$\emph{ and }$\left\vert
\phi\right\rangle $\emph{ is such that }$\left\langle \phi\right\vert
\widehat{S}^{\dag}\widehat{q}\widehat{S}\left\vert \phi\right\rangle =\gamma
$\emph{ }- e.g., when $\left\vert \left\langle q\right\vert \left.
\phi\right\rangle \right\vert ^{2}$ is Gaussian with $0$ mean - \emph{the
pointer position can be used to determine }$\gamma$ since - from eq.(\ref{9})%
\[
\left\langle \Psi\right\vert \widehat{q}\left\vert \Psi\right\rangle =\gamma.
\]
This is clearly the PPS analogue of eq.(\ref{5}) when $\left\langle
\psi\right\vert \widehat{A}\left\vert \psi\right\rangle =1$.

\section{The Weak Measurement Regime}

In this section, the exact results given by eqs.(\ref{4}) and (\ref{9}) of the
last section are used to obtain approximate results for these quantities that
are valid for PS and PPS systems when the measurements are weak. In this case
$0<\gamma\ll1$, so that to $1^{st}$ order in $\gamma$
\begin{equation}
\widehat{S}\simeq\widehat{1}-\frac{i}{\hbar}\gamma\widehat{p}\text{ \ and
}\widehat{S}^{\dag}\simeq\widehat{1}+\frac{i}{\hbar}\gamma\widehat{p}.
\label{10}%
\end{equation}
Observe that when this approximation to $\widehat{S}$ is applied to
eq.(\ref{1}), then $\widehat{1}-\widehat{A}+\widehat{A}\widehat{S}%
\simeq\widehat{1}-\frac{i}{\hbar}\gamma\widehat{A}\widehat{p}$ \ is the
$1^{st}$ order approximation to the von Neumann measurement interaction
operator. Consequently, the idempotency of $\widehat{A}$ is not relevant to
this approximation and - since weak measurement results are expressed only
through $1^{st}$ order in $\gamma$ - all of \emph{the following results
associated with measurements of PS and PPS systems in the weak measurement
regime apply for any operator} $\widehat{A}$ (projector or non-projector). In
what follows, it is assumed that the weakness conditions (inequalities (3.5)
in \cite{PCS}) are satisfied for PPS systems.

\subsection{PS Systems}

Application of approximations (\ref{10}) to eq.(\ref{4}) yields%
\begin{equation}
\left\langle \Phi\right\vert \widehat{M}\left\vert \Phi\right\rangle
\simeq\left\langle \phi\right\vert \widehat{M}\left\vert \phi\right\rangle
-\frac{i}{\hbar}\gamma\left\langle \psi\right\vert \widehat{A}\left\vert
\psi\right\rangle \left\langle \phi\right\vert \left[  \widehat{M},\widehat
{p}\right]  \left\vert \phi\right\rangle \label{11}%
\end{equation}
as the approximate mean value of the pointer observable $M$ obtained from the
weak measurement of \emph{any} observable $A$ when the PS system is in the
prepared state $\left\vert \psi\right\rangle $ at the time of measurement. It
is seen from eq.(\ref{11}) that when $\widehat{M}=\widehat{p}$, then - since
$\left[  \widehat{p},\widehat{p}\right]  =0$ - $\left\langle \Phi\right\vert
\widehat{p}\left\vert \Phi\right\rangle $ is unchanged by a weak measurement
process. As required, this is in complete agreement with eq.(\ref{4a}). Also,
when $\widehat{M}=\widehat{q}$, then - since $\left[  \widehat{q},\widehat
{p}\right]  =i\hbar$ - eq.(\ref{11}) becomes%
\[
\left\langle \Phi\right\vert \widehat{q}\left\vert \Phi\right\rangle
\simeq\left\langle \phi\right\vert \widehat{q}\left\vert \phi\right\rangle
+\gamma\left\langle \psi\right\vert \widehat{A}\left\vert \psi\right\rangle
\]
which is in agreement with eq.(\ref{5}) whenever $\left\langle \phi\right\vert
\widehat{q}\left\vert \phi\right\rangle =0$.

In general, the calculus of error propagation provides the measurement
sensitivity $\delta\left\langle \psi\right\vert \widehat{A}\left\vert
\psi\right\rangle $ which is defined to be the positive square root of%
\begin{equation}
\delta^{2}\left\langle \phi\right\vert \widehat{A}\left\vert \phi\right\rangle
\equiv\frac{\Delta_{\Phi}^{2}M}{\left\vert \frac{\partial\left\langle
\Phi\right\vert \widehat{M}\left\vert \Phi\right\rangle }{\partial\left\langle
\psi\right\vert \widehat{A}\left\vert \psi\right\rangle }\right\vert ^{2}%
}\label{11a}%
\end{equation}
as an estimate of the accuracy associated with the determination of
$\left\langle \psi\right\vert \widehat{A}\left\vert \psi\right\rangle $ from
the measurement of the mean value of pointer observable $M$. Here%
\[
\Delta_{\Phi}^{2}M=\left\langle \Phi\right\vert \widehat{M}^{\text{ }%
2}\left\vert \Phi\right\rangle -\left\langle \Phi\right\vert \widehat
{M}\left\vert \Phi\right\rangle ^{2}%
\]
is the measurement variance of $M$ relative to the final pointer state
$\left\vert \Phi\right\rangle $. In the weak measurement regime this variance
- upon application of eq.(\ref{11}) - becomes%
\[
\Delta_{\Phi}^{2}M\simeq\Delta_{\phi}^{2}M-\frac{i}{\hbar}\gamma B\left(
\widehat{M},\widehat{p}\right)  \left\langle \psi\right\vert \widehat
{A}\left\vert \psi\right\rangle ,
\]
where%
\[
B\left(  \widehat{M},\widehat{p}\right)  \equiv\left\langle \phi\right\vert
\left[  \widehat{M}^{2},\widehat{p}\right]  \left\vert \phi\right\rangle
-2\left\langle \phi\right\vert \widehat{M}\left\vert \phi\right\rangle
\left\langle \phi\right\vert \left[  \widehat{M},\widehat{p}\right]
\left\vert \phi\right\rangle
\]
and
\[
\frac{\partial\left\langle \Phi\right\vert \widehat{M}\left\vert
\Phi\right\rangle }{\partial\left\langle \psi\right\vert \widehat{A}\left\vert
\psi\right\rangle }\simeq-\frac{i}{\hbar}\gamma\left\langle \phi\right\vert
\left[  \widehat{M},\widehat{p}\right]  \left\vert \phi\right\rangle .
\]
Using these results in eq.(\ref{11a}) yields
\begin{equation}
\delta^{2}\left\langle \psi\right\vert \widehat{A}\left\vert \psi\right\rangle
\simeq\frac{\Delta_{\phi}^{2}M-\frac{i}{\hbar}\gamma B\left(  \widehat
{M},\widehat{p}\right)  \left\langle \psi\right\vert \widehat{A}\left\vert
\psi\right\rangle }{\left\vert \frac{i}{\hbar}\gamma\left\langle
\phi\right\vert \left[  \widehat{M},\widehat{p}\right]  \left\vert
\phi\right\rangle \right\vert ^{2}},\text{ \ }0<\gamma\ll1,\text{ \ }%
\widehat{M}\neq\widehat{p},\label{11b}%
\end{equation}
as the desired sensitivity approximation.

As a useful special case consider the measurement sensitivity when
$\widehat{M}=\widehat{q}$. Since $\left[  \widehat{q},\widehat{p}\right]
=i\hbar$, then $B\left(  \widehat{q},\widehat{p}\right)  =0$ and the square
root of the last equation becomes%
\[
\delta\left\langle \psi\right\vert \widehat{A}\left\vert \psi\right\rangle
\simeq\frac{\Delta_{\phi}q}{\gamma},\text{ \ }0<\gamma\ll1\text{.}%
\]
Recall from the discussion above that if $\left\langle \psi\right\vert
\widehat{A}\left\vert \psi\right\rangle $ is known and $\left\langle
\phi\right\vert \widehat{q}\left\vert \phi\right\rangle =0$, then $\gamma$ can
also be determined from the measurement of $\left\langle \Phi\right\vert
\widehat{q}\left\vert \Phi\right\rangle $. The sensitivity $\delta\gamma$
associated with this measurement is%
\begin{equation}
\delta\gamma\simeq\frac{\Delta_{\phi}q}{\left\langle \psi\right\vert
\widehat{A}\left\vert \psi\right\rangle },\text{ \ }0<\gamma\ll1, \label{11c}%
\end{equation}
which follows from the application of
\[
\frac{\partial\left\langle \Phi\right\vert \widehat{q}\left\vert
\Phi\right\rangle }{\partial\gamma}\simeq\left\langle \psi\right\vert
\widehat{A}\left\vert \psi\right\rangle
\]
to the appropriate analogue of eq.(\ref{11a}). These intuitively pleasing
results clearly show the accuracy trade-offs associated with weak measurements
of PS systems when they are used to obtain $\left\langle \psi\right\vert
\widehat{A}\left\vert \psi\right\rangle $ or $\gamma$ from measurements of the
mean pointer position. In particular, \emph{the accuracy of the determination
of }$\left\langle \psi\right\vert \widehat{A}\left\vert \psi\right\rangle $
($\gamma$)\emph{ from a measurement of }$q$ \emph{can be arbitrarily increased
only when }$\Delta_{\phi}q$ \emph{can be made arbitrarily small relative to
}$\gamma$ ($\left\langle \psi\right\vert \widehat{A}\left\vert \psi
\right\rangle $).

\subsection{PPS Systems}

The weak measurement approximation for eq.(\ref{9}) is given by%
\begin{equation}
\left\langle \Psi\right\vert \widehat{M}\left\vert \Psi\right\rangle
\simeq\left\langle \phi\right\vert \widehat{M}\left\vert \phi\right\rangle
-\frac{i}{\hbar}\gamma A_{w}^{\ast}\left\langle \phi\right\vert \left[
\widehat{M},\widehat{p}\right]  \left\vert \phi\right\rangle +2\frac{\gamma
}{\hbar}\operatorname{Im}A_{w}\cdot ccv\left(  \widehat{M},\widehat{p}\right)
,\label{12}%
\end{equation}
where $ccv\left(  \widehat{M},\widehat{p}\right)  $ is the "complex
covariance" of $\widehat{M}$ and $\widehat{p}$ relative to the initial pointer
state $\left\vert \phi\right\rangle $ defined as%
\[
ccv\left(  \widehat{M},\widehat{p}\right)  \equiv\left\langle \phi\right\vert
\widehat{M}\widehat{p}\left\vert \phi\right\rangle -\left\langle
\phi\right\vert \widehat{M}\left\vert \phi\right\rangle \left\langle
\phi\right\vert \widehat{p}\left\vert \phi\right\rangle .
\]
Since this quantity is related to the associated real valued covariance
$cov\left(  \widehat{M},\widehat{p}\right)  $ defined by \cite{JL}
\[
cov\left(  \widehat{M},\widehat{p}\right)  \equiv\frac{1}{2}\left\langle
\phi\right\vert \left\{  \widehat{M},\widehat{p}\right\}  \left\vert
\phi\right\rangle -\left\langle \phi\right\vert \widehat{M}\left\vert
\phi\right\rangle \left\langle \phi\right\vert \widehat{p}\left\vert
\phi\right\rangle
\]
according to%
\[
2cov\left(  \widehat{M},\widehat{p}\right)  =ccv\left(  \widehat{M}%
,\widehat{p}\right)  +ccv\left(  \widehat{p},\widehat{M}\right)  ,
\]
then the "complex covariance" measures - in some sense - the strength of the
"correlation" between $\widehat{M}$ and $\widehat{p}$ relative to $\left\vert
\phi\right\rangle $. Here $\left\{  \widehat{M},\widehat{p}\right\}
\equiv\widehat{M}\widehat{p}+\widehat{p}\widehat{M}$ is the anti-commutator of
$\widehat{M}$ and $\widehat{p}$.

Although written in a somewhat different form, it should be noted that
eq.(\ref{12}) agrees precisely with that of eq.(17) in \cite{Jo}.
Consequently, the results also agree for the special cases $\widehat
{M}=\widehat{p}$ and $\widehat{M}=\widehat{q}$. In particular, since
$ccv\left(  \widehat{p},\widehat{p}\right)  =\Delta_{\phi}^{2}p$ and $\left[
\widehat{p},\widehat{p}\right]  =0$, then - in complete agreement with
\cite{Jo} -%
\begin{equation}
\left\langle \Psi\right\vert \widehat{p}\left\vert \Psi\right\rangle
\simeq\left\langle \phi\right\vert \widehat{p}\left\vert \phi\right\rangle
+2\frac{\gamma}{\hbar}\operatorname{Im}A_{w}\cdot\Delta_{\phi}^{2}%
p.\label{12a}%
\end{equation}
Also, it follows that for a pointer of mass $m$
\begin{equation}
2ccv\left(  \widehat{q},\widehat{p}\right)  =\left[  \widehat{q},\widehat
{p}\right]  +m\frac{d}{dt}\left(  \Delta_{\phi}^{2}q\right)  .\label{13}%
\end{equation}
Using this and $\left[  \widehat{q},\widehat{p}\right]  =i\hbar$ in
eq.(\ref{12}) gives%
\begin{equation}
\left\langle \Psi\right\vert \widehat{q}\left\vert \Psi\right\rangle
\simeq\left\langle \phi\right\vert \widehat{q}\left\vert \phi\right\rangle
+\gamma\operatorname{Re}A_{w}+m\frac{\gamma}{\hbar}\operatorname{Im}A_{w}%
\cdot\frac{d}{dt}\left(  \Delta_{\phi}^{2}q\right)  \label{13a}%
\end{equation}
which is also in complete agreement with \cite{Jo}.

The compact form of eq.(\ref{12}) is convenient for identifying the effect of
post-selection upon the mean value of $M$ when the measurements are weak.
Comparison with eq.(\ref{11}) reveals that - in addition to replacing
$\left\langle \psi\right\vert \widehat{A}\left\vert \psi\right\rangle $ with
$A_{w}^{\ast}$ in the second term - \emph{post-selection} also \emph{induces
the peculiar third term containing the "complex covariance" factor that
"correlates" the observable }$M$\emph{ with the pointer momentum }$p$\emph{ -
even when }$\left[  \widehat{M},\widehat{p}\right]  =0$. Since this term
depends upon $\operatorname{Im}A_{w}$, this "correlation" only exists when
$\operatorname{Im}A_{w}\neq0$. Thus, if $\operatorname{Im}A_{w}\neq0$,
\emph{this "correlation" is manifested as : (i) pointer momentum variance when
}$\widehat{M}=\widehat{p}$\emph{ ; and (ii) the sum of the quantum dynamical
term }$\left[  \widehat{q},\widehat{p}\right]  $\emph{ and the rate of change
of pointer position variance just prior to measurement time (for massive
pointers) when }$\widehat{M}=\widehat{q}$. This is consistent with the
observation in \cite{Jo} that the second term in eq.(\ref{12a}) is "an
artifact of post-selection rather than a quantum dynamical effect" whereas the
translation of the mean pointer position in eq.(\ref{13a}) is quantum
dynamical in nature. Note from eq.(\ref{13}) - however - that
\emph{post-selection } \emph{"correlates" }$q$\emph{ and }$p$\emph{ through
the quantum dynamical term }$\left[  \widehat{q},\widehat{p}\right]  $
\emph{in }$ccv\left(  \widehat{q},\widehat{p}\right)  $\emph{ even when
}$\frac{d}{dt}\left(  \Delta^{2}q\right)  =0$. Clearly, if $A_{w}$ is real
valued, then the third "complex covariance" term in eq.(\ref{12}) vanishes and
eqs.(\ref{11}) and (\ref{12}) share a similar two term form.

The measurement sensitivity $\delta\operatorname{Re}A_{w}$ follows from the
ratio%
\[
\delta^{2}\operatorname{Re}A_{w}\equiv\frac{\Delta_{\Psi}^{2}M}{\left\vert
\frac{\partial\left\langle \Psi\right\vert \widehat{M}\left\vert
\Psi\right\rangle }{\partial\operatorname{Re}A_{w}}\right\vert ^{2}}%
\]
where - from eq.(\ref{12}) -%
\[
\Delta_{\Psi}^{2}M\simeq\Delta_{\phi}^{2}M-\frac{i}{\hbar}\gamma B\left(
\widehat{M},\widehat{p}\right)  \operatorname{Re}A_{w}+\frac{\gamma}{\hbar
}\left[  C\left(  \widehat{M},\widehat{p}\right)  -2\left\langle
\phi\right\vert \widehat{p}\left\vert \phi\right\rangle \left(  \Delta_{\phi
}^{2}M-\left\langle \phi\right\vert \widehat{M}\left\vert \phi\right\rangle
^{2}\right)  \right]  \operatorname{Im}A_{w},
\]
with%
\[
C\left(  \widehat{M},\widehat{p}\right)  \equiv\left\langle \phi\right\vert
\left\{  \widehat{M}^{2},\widehat{p}\right\}  \left\vert \phi\right\rangle
-2\left\langle \phi\right\vert \widehat{M}\left\vert \phi\right\rangle
\left\langle \phi\right\vert \left\{  \widehat{M},\widehat{p}\right\}
\left\vert \phi\right\rangle
\]
and%
\[
\frac{\partial\left\langle \Psi\right\vert \widehat{M}\left\vert
\Psi\right\rangle }{\partial\operatorname{Re}A_{w}}\simeq-\frac{i}{\hbar
}\gamma\left\langle \phi\right\vert \left[  \widehat{M},\widehat{p}\right]
\left\vert \phi\right\rangle .
\]
It is interesting to note that $C\left(  \widehat{M},\widehat{p}\right)  $
$\rightarrow$ $B\left(  \widehat{M},\widehat{p}\right)  $ when the
anti-commutators in $C\left(  \widehat{M},\widehat{p}\right)  $ are replaced
with commutators.

Comparison of $\delta^{2}\operatorname{Re}A_{w}$with eq.(\ref{11b}) shows that
if $A_{w}$ is real valued, then the accuracy does not depend upon $C\left(
\widehat{M},\widehat{p}\right)  $ and the difference $\left(  \Delta_{\phi
}^{2}M-\left\langle \phi\right\vert \widehat{M}\left\vert \phi\right\rangle
^{2}\right)  $. Thus, in this case - except for the $\left\langle
\psi\right\vert \widehat{A}\left\vert \psi\right\rangle $ and
$\operatorname{Re}A_{w}$ factors - $\delta^{2}\left\langle \psi\right\vert
\widehat{A}\left\vert \psi\right\rangle $ and $\delta^{2}\operatorname{Re}%
A_{w}$ have the same form (in fact, if $\left\vert \psi_{i}\right\rangle
=\left\vert \psi_{f}\right\rangle =\left\vert \psi\right\rangle $, then
$\delta\left\langle \psi\right\vert \widehat{A}\left\vert \psi\right\rangle
=\delta\operatorname{Re}A_{w}$). It is also clear from this that \emph{if
}$A_{w}$\emph{ is complex valued, then }$\operatorname{Im}A_{w}$\emph{ effects
the accuracy of }$\operatorname{Re}A_{w}$\emph{ when it is determined from a
measurement of the mean value of }$M$.

When $\widehat{M}=\widehat{q}$ it has already been noted elsewhere that
$\operatorname{Im}A_{w}$ has an impact upon the pointer spatial distribution
profile \cite{A2} and consequently can change the size of the associated PPS
ensemble \cite{PDX}. In addition to this - the above results also show the
effect of $\operatorname{Im}A_{w}$ upon the accuracy associated with mean
pointer position measurements of PPS systems. Specifically, when $\widehat
{M}=\widehat{q}$, then $B\left(  \widehat{q},\widehat{p}\right)  =0$ and
$C\left(  \widehat{q},\widehat{p}\right)  \neq0$ so that%
\[
\delta^{2}\operatorname{Re}A_{w}\simeq\delta^{2}\left\langle \psi\right\vert
\widehat{A}\left\vert \psi\right\rangle +\left(  \frac{1}{\gamma\hbar}\right)
\left[  C\left(  \widehat{q},\widehat{p}\right)  -2\left\langle \phi
\right\vert \widehat{p}\left\vert \phi\right\rangle \left(  \Delta_{\phi}%
^{2}q-\left\langle \phi\right\vert \widehat{q}\left\vert \phi\right\rangle
^{2}\right)  \right]  \operatorname{Im}A_{w},\text{ \ }0<\gamma\ll1.
\]
Thus, \emph{the effect of a complex valued }$A_{w}$\emph{ is to increase
}(\emph{decrease})\emph{ the accuracy of }$\operatorname{Re}A_{w}$\emph{ if it
is determined from a measurement of the mean value of }$q$\emph{ whenever}%
\[
\left[  C\left(  \widehat{q},\widehat{p}\right)  -2\left\langle \phi
\right\vert \widehat{p}\left\vert \phi\right\rangle \left(  \Delta_{\phi}%
^{2}q-\left\langle \phi\right\vert \widehat{q}\left\vert \phi\right\rangle
^{2}\right)  \right]  \operatorname{Im}A_{w}<0\text{ }\left(  >0\right)  .
\]
It is important to note that it may be possible to exploit this effect to
increase the accuracy of the determination of $\operatorname{Re}A_{w}$.
\emph{If }$A_{w}$\emph{ is real valued, then}%
\[
\delta A_{w}=\delta\left\langle \psi\right\vert \widehat{A}\left\vert
\psi\right\rangle .
\]

For the sake of completeness, consider the case mentioned above where it was
noted that $\gamma$ can be determined from a measurement of $\left\langle
\Psi\right\vert \widehat{q}\left\vert \Psi\right\rangle $ when $\left\langle
\phi\right\vert \widehat{q}\left\vert \phi\right\rangle =0$ and $A_{w}=1$.
Since%
\[
\Delta_{\Psi}^{2}q\simeq\Delta_{\phi}^{2}q
\]
and%
\[
\frac{\partial\left\langle \Psi\right\vert \widehat{q}\left\vert
\Psi\right\rangle }{\partial\gamma}\simeq1,
\]
then%
\[
\delta\gamma\simeq\Delta_{\phi}q
\]
which is clearly the analogue of eq.(\ref{11c}) when $\left\langle
\psi\right\vert \widehat{A}\left\vert \psi\right\rangle =1$.

\section{Concluding Remarks}

The idempotent property of projection operators has been used to provide for
any measurement interaction strength exact simple expressions for both state
vectors and arbitrary pointer observable mean values that are associated with
projector measurements of PS and PPS quantum systems. These results
demonstrate that : (i) the idempotency of the projector precludes the
existence of interference cross terms in the distribution profiles and in the
exact expressions for pointer observable mean values for PS systems ; (ii)
post-selection nullifies the effect of projector idempotency in projector
measurements of PPS systems so that interference cross-terms appear in the
distribution profiles (thereby providing an observable distinction between PS
and PPS systems) and in the exact expressions for pointer observable mean
values ; (iii) post-selection induces a Pancharatnam phase into the exact
states for PPS systems ; (iv) regardless of the strength of the interaction
both the exact state and the exact expression for the mean value of a pointer
observable for PPS systems depends upon the weak value of the projector ; (v)
whereas pointer momentum is a constant of the motion for projector
measurements of PS systems - it is not a constant of the motion for projector
measurements of PPS systems ; (vi) measurement interaction strengths and mean
values for projectors can both be straightforwardly determined from mean
pointer position measurements of PS systems when the mean pre-measurement
position of the pointer is zero ; and (vii) only measurement interaction
strengths can be straightforwardly determined from mean pointer position
measurements of PPS systems when the weak value of the projector is unity.

When applied to the weak measurement regime these results demonstrate that :
(i) the exact pointer observable mean values yield approximate expressions
which are valid for any operator (projector or non-projector) ; (ii) the
approximate expression for the mean value of an arbitrary pointer observable
for PPS systems agrees exactly with \cite{Jo} ; (iii) complex valued weak
values "correlate" the pointer observable for a PPS system with the pointer
momentum - even when the pointer observable and momentum commute ; (iv) the
accuracy associated with determining the mean value of an operator from a
measurement of the mean pointer position for a PS system can be made
arbitrarily small only when the pointer's pre-measurement position uncertainty
can be made arbitrarily small relative to the strength of the measurement
interaction ; (v) the accuracy associated with determining the real part of
the weak value of an operator from a measurement of the mean pointer position
for PPS systems is effected by the imaginary part of the weak value (it may be
possible to exploit this to increase the measurement accuracy of the real part
of a complex weak value) ; and (vi) when the weak value of an operator is real
valued the accuracy associated with determining its weak value from a mean
pointer position measurement of a PPS system is precisely the same as the
accuracy associated with determining its mean value from a mean pointer
position measurement of a PS system.

\end{document}